\documentstyle[aps,12pt,epsf]{revtex}

\begin{document}
\draft
\author{O. B. Zaslavskii}
\address{Department of Physics, Kharkov State University, Svobody Sq.4, Kharkov\\
310077, Ukraine\\
E-mail: aptm@kharkov.ua}
\title{Exactly solvable models of two-dimensional dilaton gravity and quantum
eternal black holes}
\maketitle

\begin{abstract}
New approach to exact solvability of dilaton gravity theories is suggested
which appeals directly to structure of field equations. It is shown that
black holes regular at the horizon are static and their metric is found
explicitly. If a metric possesses singularities the whole spacetime can be
divided into different sheets with one horizon on each sheet between
neighboring singularities with a finite value of dilaton field (addition
horizons may arise at infinite value of it), neighboring sheets being glued
along the singularity. The position of singularities coincide with the
values of dilaton in solutions with a constant dilaton field. Quantum
corrections to the Hawking temperature vanish. For a wide subset of these
models the relationship between the total energy and the total entropy of
the quantum finite size system is the same as in the classical limit. For
another subset the metric itself does not acquire quantum corrections. The
present paper generalizes Solodukhin's results on the RST model in that
instead of a particular model we deal with whole classes of them. Apart from
this, the found models exhibit some qualitatively new properties which are
absent in the RST model. The most important one is that there exist quantum
black holes with geometry regular everywhere including infinity.
\end{abstract}


\section{introduction}

Two-dimensional theories of gravity count on better understanding the role
of quantum effects in black hole physics in a more realistic
four-dimensional case. In particular, new insight was gained due to reducing
the problem of black hole evaporation to solving differential equations of
the semiclassical theory \cite{callan}. However, these equations remain too
complicated in the sense that exact solutions cannot be found. This obstacle
was overcome in the approach based on modifying the form of the
semiclassical action in such a way that solvability is restored. The
well-known example is the Russo-Susskind-Thorlacius (RST) model \cite{rst}.
In particular, this enabled one to give exhausting analysis of either
geometry or thermodynamics of eternal black holes in the framework of RST
dilaton gravity \cite{solod}.

Examples of exactly solvable theories of dilaton gravity were discussed in 
\cite{bil}, \cite{alw}. As was shown by Kazama, Satoh and Tsuichiya (KST),
these models as well as the RST one can be found as particular cases of the
unified approach \cite{kaz} based on the symmetries of the nonlinear sigma
model. A number of other exactly solvable models were suggested later \cite
{rob}, \cite{fub}, \cite{bose}, \cite{cruz}.

The aim of the present paper is two-fold. First, we suggest new approach to
finding criteria for exact solvability and establish its equivalence to that
of \cite{kaz}. We also show that our approach encompasses all particular
models mentioned above. Second, we solve the field equations in a closed
form and study general properties of the found spacetimes. In this point we
generalize the Solodukhin's results for RST eternal black holes. The
essential feature of models considered below consists in that we specify not
the explicit dependence of the action coefficients on dilaton but, rather,
relationship between these coefficients. This means that instead of one or
several particular exactly solvable models we deal at once with the whole
class of them.

The paper is organized as follows. In Sec. II we describe our approach and
derive the basic equation which selects exactly-solvable models of
two-dimensional dilaton gravity among all possible forms. We find the
relation between action coefficients and demonstrate how it enables one to
cast the theory into the Liouville form. We also briefly show that the this
relation is equivalent to that obtained in \cite{kaz} in a quite different
approach. Further, we prove that all black holes with a regular geometry are
static and find their metric explicitly. In Sec. III we discuss properties
of found solutions - spacetime structure, thermodynamics, the existence of a
special class of solutions with a constant dilaton field value, etc. In Sec.
IV we show that all exactly solvable models mentioned above fall in the
found class. In Sec. V we summarize the main features of our approach and
properties of obtained solutions and outline some perspectives for future
researches.

\section{basic equations}

\subsection{Conditions of exact solvability}

Let us consider the system described by the action 
\begin{equation}
I=I_{0}+I_{PL}  \label{1}
\end{equation}
where 
\begin{equation}
I_{0}=\frac{1}{2\pi }\int_{M}d^{2}x\sqrt{-g}[F(\phi )R+V(\phi )(\nabla \phi
)^{2}+U(\phi )]+\frac{1}{\pi }\int_{\partial M}dskF(\phi )  \label{2}
\end{equation}
Here the boundary term with the second fundamental form $k$ makes the
variational problem self-consistent, $ds$ is the line element along the
boundary $\partial M$ of the manifold $M.$

$I_{PL\text{ }}$is the Polyakov-Liouville action \cite{polyakov}
incorporating effects of Hawking radiation and its back reaction on the
black hole metric for a multiplet of N scalar fields. It is convenient to
write it down in the form \cite{solod}, \cite{israel} 
\begin{equation}
I_{PL}=-\frac{\kappa }{2\pi }\int_{M}d^{2}x\sqrt{-g}[\frac{(\nabla \psi )^{2}%
}{2}+\psi R]-\frac{\kappa }{\pi }\int_{\partial M}\psi kds  \label{3}
\end{equation}
The function $\psi $ obeys the equation 
\begin{equation}
\Box \psi =R  \label{4}
\end{equation}
where $\Box =\nabla _{\mu }\nabla ^{\mu }$, $\kappa =N/24$ is the quantum
coupling parameter.

We imply that all coefficients $F,V,U$ in (\ref{2}) may, in general, contain
terms with $\kappa .$ For instance, in the RST model $F=e^{-2\phi }-\kappa
\phi .$

Varying the action with respect to metric gives us $(T_{\mu \nu }=2\frac{%
\delta I}{\delta g^{\mu \nu }})$: 
\begin{equation}
T_{\mu \nu }\equiv T_{\mu \nu }^{(0)}+T_{\mu \nu }^{(PL)}=0  \label{6}
\end{equation}
where 
\begin{equation}
T_{\mu \nu }^{(0)}=\frac{1}{2\pi }\{2(g_{\mu \nu }\Box F-\nabla _{\mu
}\nabla _{\nu }F)-Ug_{\mu \nu }+2V\nabla _{\mu }\phi \nabla _{\nu }\phi
-g_{\mu \nu }V(\nabla \phi )^{2}\}\text{,}  \label{7}
\end{equation}
\begin{equation}
T_{\mu \nu }^{(PL)}=-\frac{\kappa }{2\pi }\{\partial _{\mu }\psi \partial
_{\nu }\psi -2\nabla _{\mu }\nabla _{\nu }\psi +g_{\mu \nu }[2R-\frac{1}{2}%
(\nabla \psi )^{2}]\}  \label{8}
\end{equation}

Variation of the action with respect to $\phi $ gives rise to the equation 
\begin{equation}
RF^{^{\prime }}+U^{\prime }=2V\Box \phi +V^{\prime }(\nabla \phi )^{2}\text{,%
}  \label{9}
\end{equation}
where prime denotes derivative with respect to $\phi $.

In general, eqs. (\ref{4}) - (\ref{9}) cannot be solved exactly. In
particular, the auxiliary function $\psi $ is a rather complicated nonlocal
functional of dilaton field $\phi $ and its derivatives, the explicit form
of which cannot be found. The key idea for finding exactly solvable models
consists in two assumptions. First, we select among all variety of models
such a subset for which the above equations admit the solution $\psi =\psi
(\phi )$, i.e. a local connection between $\psi $ and $\phi .$ As was shown
by Solodukhin \cite{solod}, for eternal black holes in the RST model $\psi
=-2\phi .$ Now the dependence in question can be, generally speaking,
nonlinear. Second, we impose such a constraint on the coefficients of the
action that enables us to single out the terms with derivatives of $\phi $
with respect to coordinates. As a result, instead of solving the original
(rather complicated equations) it is sufficient to ensure the cancellation
of the coefficients at $\Box \phi $ and $(\nabla \phi )^{2}$ which
themselves already do not contain derivatives with respect to coordinates.
It turns out that such a constraint is not very tight and includes a large
variety of models which may be of physical interest.

Inasmuch as the auxiliary function $\psi $ can be expressed in terms of $%
\phi $ directly, the action $I_{0\text{ }}$and the Polyakov-Liouville action
are combined in such a way that field equations (\ref{6}) - (\ref{8}) can be
formally obtained from the action $I_{0}$ only but with the ''renormalized''
coefficients which receive some shifts: $F\rightarrow \tilde{F},V\rightarrow 
\tilde{V}$ where 
\begin{equation}
\tilde{F}=F-\kappa \psi \text{,}  \label{10}
\end{equation}
\begin{equation}
\tilde{V}=V-\frac{\kappa }{2}\psi ^{\prime 2}  \label{11}
\end{equation}

Taking the trace of eq. (\ref{6}) we get the relation 
\begin{equation}
U=\Box \tilde{F}  \label{12}
\end{equation}
Then eq. (\ref{9}), with eqs. (\ref{4}), (\ref{10}) - (\ref{12}) taken into
account, reads 
\begin{eqnarray}
&&A_{1}\Box \phi +A_{2}(\nabla \phi )^{2}=0\text{,}  \nonumber \\
&&A_{1}=(u-\kappa \omega )\psi ^{\prime }+\omega u-2V\text{,}  \label{13} \\
&&A_{2}=(u-\kappa \omega )\psi ^{\prime \prime }+\omega u^{\prime
}-V^{\prime }\text{,}  \nonumber
\end{eqnarray}
where $\omega \equiv U^{\prime }/U,$ $u\equiv F^{\prime }.$ For arbitrary
coefficients $A_{1}(\phi )$, $A_{2}(\phi )$ eq.(\ref{13}) cannot be solved
exactly. This can be done, however, if we demand that both coefficients in
eq.(\ref{13}) turn into zero. Such a demand represents the {\it sufficient }%
condition for eq.(\ref{13}) to be satisfied. Then it follows from eq. $%
A_{1}=0$ that 
\begin{equation}
\psi ^{\prime }=\frac{2V-\omega u}{u-\kappa \omega }\text{,}  \label{14}
\end{equation}
which enables us to find at once $\psi $ in terms of known functions $u$, $V$%
, $\omega $ by direct integration. Demanding that both equations $A_{1}=0$
and $A_{2}=0$ be consistent with each other, we differentiate eq. $A_{1}=0$
and compare the result with $A_{2}=0$. Then we have 
\begin{equation}
u^{\prime }(2V-\omega u)+u(u\omega ^{\prime }-V^{\prime })+\kappa (\omega
V^{\prime }-2V\omega ^{\prime })=0  \label{15}
\end{equation}

Thus, we made some selections among all possible models. Eq.(\ref{15}) is
the only constraint on the relationship between these three coefficients
that leaves enough freedom in choosing a model. Both eqs.(\ref{14}), (\ref
{15}) represent direct consequences of our assumptions $\psi =\psi (\phi )$, 
$A_{1}=A_{2}=0$ and do not hold true in an arbitrary model. In particular,
for the original form of the action in the string-inspired gravity \cite
{string} $F=e^{-2\phi },V=4e^{-2\phi },\omega =-2$ these equation are
satisfied in the zero order in $\kappa $ only. However, for the RST action 
\begin{equation}
F=e^{-2\phi }-\kappa \phi ,V=4e^{-2\phi },\omega =-2  \label{model}
\end{equation}
and this model does obey eq.(\ref{15}). Then the integration of (\ref{14})
leads to the relationship $\psi =-2\phi +const$ that agrees with \cite{solod}%
.

It follows from eq. (\ref{15}) that the function $\omega $ expressed in
terms of $V$ and $u$ reads 
\begin{equation}
\omega =\frac{u-D\sqrt{u^{2}-2V\kappa }}{\kappa }  \label{w}
\end{equation}
Reverting this formula, we have 
\begin{equation}
V=\omega (u-\frac{\kappa \omega }{2})+C(u-\kappa \omega )^{2}  \label{v}
\end{equation}
where $C=(2\kappa )^{-1}(1-D^{-2})$. We must assume that the constant $%
D\rightarrow 1$ when $\kappa \rightarrow 0$ in order to have the
well-defined classical limit. In what follows we mainly restrict ourselves
to the simplest case $C\equiv 0$, $D\equiv 1$, when 
\begin{equation}
\omega =\frac{u-\sqrt{u^{2}-2V\kappa }}{\kappa }\text{, }V=\omega (u-\frac{%
\kappa \omega }{2})  \label{d=1}
\end{equation}

\subsection{Comparison with KST action}

In eq. (3.18) of Ref.\cite{kaz} the following relation between different
action coefficients was obtained: 
\begin{equation}
a_{,q}(1-\frac{\kappa }{2}W_{,q})+\kappa W_{,qq}(a+\frac{1}{2}W_{,q})=0
\label{k}
\end{equation}
where (in our notations) $W=\int d\phi \omega $ and by definition $%
V=aF^{\prime 2}+F^{\prime }\omega \equiv au^{2}+u\omega $. Substituting
these expressions into (\ref{k}) we obtain after simple but rather lengthy
calculation the equation which coincides with our eq. (\ref{15}) exactly. It
was pointed out in \cite{kaz} that results of \cite{bil}, \cite{alw} are
contained in the general formula following from (\ref{k}). In our scheme the
models of Ref. \cite{alw} (which include those of \cite{bil} as particular
cases) follow from (\ref{d=1}) if one writes down $V=4e^{-2\phi }[1+h(\phi
)] $, $u=-2e^{-2\phi }[1+\tilde{h}]$.

KST approach has the advantage of elucidating the hidden symmetry of the
action in terms of a nonlinear $\sigma $ model. On the other hand, the
present approach, in our view, is much simpler in that it operates directly
with the original action coefficients and demonstrates explicitly the origin
of solvability as the cancellation of coefficients at $(\nabla \phi )^{2}$
and $\Box \phi $ in (\ref{13}). It also enables us to establish the
staticity of black hole solutions (in the absence of external origins and
matter fields) and find their explicit form. It is this issue that we now
turn to.

\subsection{Form of the metric}

Let us return to general formulae without specifying the gauge. With eq. (%
\ref{12}) taken into account, the field equations (\ref{6}) - (\ref{8}) can
be rewritten in the form 
\begin{equation}
\lbrack \xi _{1}\Box \phi +\xi _{2}(\nabla \phi )^{2}]g_{\mu \nu }=2(\xi
_{1}\nabla _{\mu }\nabla _{\nu }\phi +\xi _{2}\nabla _{\mu }\phi \nabla
_{\nu }\phi )  \label{16}
\end{equation}
where $\xi _{1}=\tilde{F}^{\prime }$, $\xi _{2}=\tilde{F}^{\prime \prime }-%
\tilde{V}$. Let us multiply this equation by the factor $\eta $ chosen in
such a way that $(\xi _{1}\eta )^{\prime }=\xi _{2}\eta $ whence $\eta =\exp
(-\int d\phi \tilde{V}/\tilde{F}^{\prime })$. Then eq. (\ref{16}) turns into 
\begin{equation}
g_{\mu \nu }\Box \mu =2\nabla _{\mu }\nabla _{\nu }\mu  \label{17}
\end{equation}
where by definition $\mu ^{\prime }=\xi _{1}\eta .$ This equation takes the
same form as eq.(2.24) from \cite{solod} and entails the same conclusions
about properties of the geometry. It is convenient to choose the space-like
coordinate as $x=\mu (\phi )/B$ where $B$ is some constant. Then it follows
from eq.(\ref{17}) that the metric takes the Schwarzschild-like form and is
static: 
\begin{equation}
ds^{2}=-gdt^{2}+g^{-1}dx^{2}  \label{18}
\end{equation}
In this gauge $R=-\frac{d^{2}g}{dx^{2}}$ . Substituting it into eq.(\ref{4})
we get after integration: 
\begin{equation}
g=A\int_{\phi _{h}}^{\phi }d\tilde{\phi}\frac{\partial \mu }{\partial \tilde{%
\phi}}e^{\psi (\tilde{\phi})-\psi (\phi )}  \label{19}
\end{equation}
\begin{eqnarray}
x &=&B^{-1}\mu \text{, }\mu =\int^{\phi }d\tilde{\phi}\frac{\partial \tilde{F%
}}{\partial \tilde{\phi}}e^{\int^{\tilde{\phi}}d\phi ^{\prime }\alpha (\phi
^{\prime })}\text{,}  \nonumber \\
\alpha &=&-\frac{\tilde{V}}{\tilde{F}^{\prime }}\text{,}  \label{20}
\end{eqnarray}
where $A$ is a constant. Here it is supposed that a spacetime has an event
horizon at $x=x_{h}$, $\phi =\phi _{h}$.

Now we will see that the obtained expressions for the metric can be
simplified further. Substituting eq. (\ref{v}) into (\ref{19}), (\ref{20})
we find after some rearrangement that $\alpha =-[\omega +C(u-\kappa \omega
)] $ and 
\begin{eqnarray}
g &=&aC^{-1}(e^{CH}-e^{CH_{h}})e^{-\psi }\text{, }a\equiv AC(1-2\kappa C)%
\text{, }\psi =(1-2\kappa C)\int d\phi \omega +2CF\text{, }  \label{c} \\
x &=&B^{-1}\mu \text{, }\mu ^{\prime }=(1-2\kappa C)H^{\prime }\exp
[-CF-(1-\kappa C)\int d\phi \omega ]\text{, }H\equiv F-\kappa \int d\phi
\omega  \nonumber
\end{eqnarray}
It would seem that eqs. (\ref{c}) do not represent the full solution of the
problem since the functions under discussion should obey one more equation -
eq. (\ref{12}). It is remarkable, however, that, as follows from the
substitution of (\ref{c}) into (\ref{12}), this equation is satisfied
provided the constants entering formulae for the metric obey the relation 
\begin{equation}
aB^{2}e^{CH_{h}}(1-2\kappa C)=4\lambda ^{2}  \label{const}
\end{equation}
where $U=4\lambda ^{2}\exp (\int d\phi \omega )$ and the limit of
integration in the integral $\int^{\phi }d\phi \omega $ is chosen equally in
all formulae above.

In the case $C=0$ this gives us 
\begin{equation}
g=(\tilde{F}-\tilde{F}_{h})\exp [-\int d\tilde{\phi}\omega ]\text{, }\mu
^{\prime }=\tilde{F}^{\prime }e^{-\psi }\text{, }\tilde{F}=H\text{, }\psi
=\int \omega d\phi \text{, }B^{2}=4\lambda ^{2}  \label{g}
\end{equation}
where we put $a=1$ for definiteness. For the RST model (\ref{model}) we
obtain in accordance with \cite{solod} $\mu =\phi -\frac{\kappa }{4}e^{2\phi
}$, $B=-2\lambda $.

In the above formulae it was tacitly assumed that the function $\psi $ as
well as the metric itself is regular at the horizon (when the temperature is
equal to its Hawking value). As explained in detail in Ref. \cite{solod}
(Sec. 2B), such a choice of boundary conditions corresponds to a black hole
in equilibrium with Hawking radiation but does not describe formation of a
hole from a flat space due to incoming matter.

\subsection{Liouville theory}

The suggested approach enables one to cast the original action into the
Liouville form in a very simple way. In the conformal gauge 
\begin{equation}
ds^{2}=-e^{2\rho }dx^{+}dx^{-}  \label{conf}
\end{equation}
Bearing in mind that in this gauge the Polyakov action \cite{polyakov} reads 
$I_{PL}=-\frac{2\kappa }{\pi }\int d^{2}x\partial _{+}\rho \partial _{-}\rho
=\frac{\kappa }{\pi }\int d^{2}x\sqrt{g}(\nabla \rho )^{2}$ and integrating
in (\ref{2}) the term with curvature by parts we have (we omit now all
boundary terms) 
\begin{equation}
I=\frac{1}{2\pi }\int d^{2}x\sqrt{g}[V(\nabla \phi )^{2}+2\nabla \rho \nabla
F+2\kappa (\nabla \rho )^{2}+4\lambda ^{2}e^{\eta }]\text{,}  \label{i}
\end{equation}
where by definition $\eta =\int d\phi \omega $. Let now the potential $V$
obey the condition (\ref{v}). Then the action can be rewritten in the form 
\begin{equation}
I=\frac{1}{2\pi }\int d^{2}x\sqrt{g}[\nabla \eta \nabla F-\frac{\kappa }{2}%
(\nabla \eta )^{2}+C(\nabla H)^{2}+2\nabla \rho \nabla F+2\kappa (\nabla
\rho )^{2}+4\lambda ^{2}e^{\eta }]  \label{i2}
\end{equation}
where $H=F-\kappa \eta $. Now introduce new fields $\Omega $ and $\chi $
instead of $\phi $ and $\rho $ according to $H=2\kappa \Omega $, $\eta (\phi
)=2(\chi -\Omega -\rho )$. Then after simple rearrangement we obtain 
\begin{equation}
I=\frac{1}{\pi }\int d^{2}x\sqrt{g}\{\kappa [(\nabla \chi )^{2}-(\nabla
\Omega )^{2}(1-2C\kappa )]+2\lambda ^{2}e^{2(\chi -\Omega -\rho )}\}
\label{il}
\end{equation}
Using the explicit formula for the conformal gauge (\ref{conf}) we obtain
the action 
\begin{equation}
I=\frac{1}{\pi }\int d^{2}x\{2\kappa [(\partial _{+}\Omega \partial
_{-}\Omega (1-2C\kappa )-\partial _{+}\chi \partial _{-}\chi ]+\lambda
^{2}e^{2(\chi -\Omega )}\}  \label{li}
\end{equation}
which in the case $C=0$ takes the familiar form. Meanwhile, it is
instructive to rederive in the conformal gauge the obtained exact solutions
with an arbitrary $C\neq 0$. In this gauge the equations of motion which
follow from (\ref{li}) read 
\begin{eqnarray}
2\kappa (1-2C\kappa )\partial _{+}\partial _{-}\Omega &=&-\lambda
^{2}e^{2(\chi -\Omega )}\text{,}  \label{motion} \\
2\kappa \partial _{+}\partial _{-}\chi &=&-\lambda ^{2}e^{2(\chi -\Omega )} 
\nonumber
\end{eqnarray}
As usual, they should be supplemented by the constraint equations $%
T_{++}=T_{--}=0$. The expressions for the classical part of $T_{++\text{ }}$
and $T_{--}$ follow directly from (\ref{7}). The formula for the quantum
contribution can be obtained from the conservation law and conformal anomaly
and has the form 
\begin{equation}
T_{\pm \pm }^{(PL)}=\frac{2\kappa }{\pi }[(\partial _{\pm }\rho
)^{2}-\partial _{\pm }^{2}\rho +t_{\pm }]  \label{stress}
\end{equation}

The functions $t_{\pm }$ depend on the choice of boundary conditions. For
the Hartle-Hawking state we deal with $t_{\pm }=0$. With the potential $V$
from (\ref{v}) the constraint equations take the form 
\begin{equation}
-\partial _{\pm }^{2}F+2\partial _{\pm }\rho \partial _{\pm }\rho +(\partial
_{\pm }\phi )^{2}V+2\kappa [(\partial _{\pm }\rho )^{2}-\partial _{\pm
}^{2}\rho ]=0  \label{mot+}
\end{equation}
As seen from eq.(\ref{motion}), it is convenient to use the gauge in which $%
\chi =\Omega (1-2C\kappa )$, so $\chi -\Omega =-CH$. Then after some algebra
we find that the factor $(1-2C\kappa )$ in a remarkable way is singled out
in constraints and we arrive at equations obtained from (\ref{mot+}) and (%
\ref{motion}) 
\begin{eqnarray}
\partial _{\pm }^{2}H+C(\partial _{\pm }H)^{2} &=&0\text{,}  \label{eqh} \\
(1-2\kappa C)\partial _{+}\partial _{-}H &=&-\lambda ^{2}e^{-2CH}  \nonumber
\end{eqnarray}
It is convenient to make the substitution $z=e^{C(H-H_{h})}$ where $H_{h}$
is some constant. Then the first equation in (\ref{eqh}) turns into $%
\partial _{\pm }^{2}z=0$ whence $z=bx^{+}x^{-}+d$ (linear terms can always
be removed by shifts in coordinates). By substitution in the second eq.(\ref
{eqh}) we obtain the restriction $bd=-\tilde{\lambda}^{2}$ where $\tilde{%
\lambda}^{2}=\lambda ^{2}C(1-2\kappa C)^{-1}e^{-2CH_{h}}$. Let us choose $%
d=1 $, $b=-\tilde{\lambda}^{2}$ and make transformation to new coordinates
according to $\tilde{\lambda}x^{+}=e^{\tilde{\lambda}\sigma ^{+}}$, $-\tilde{%
\lambda}x^{-}=-e^{-\tilde{\lambda}x^{-}}$. Then the metric (\ref{conf})
turns into 
\begin{equation}
ds^{2}=-gd\sigma ^{+}d\sigma ^{-}  \label{sigma}
\end{equation}
where 
\begin{equation}
g=e^{-\psi }\frac{(e^{CH}-e^{CH_{h}})}{e^{CH_{h}}}\text{, }\psi =(1-2\kappa
C)\int d\phi \omega +2CF  \label{met}
\end{equation}
To rewrite the metric in the Schwarzschild gauge (\ref{18}), let us
introduce coordinates $\sigma ^{\pm }=t\pm \sigma $ and $d\sigma =dxg^{-1}$.
Then 
\begin{equation}
x^{\prime }(\phi )=\frac{CH^{\prime }}{2\tilde{\lambda}}e^{-CH_{h}-CF-(1-%
\kappa C)\int d\phi \omega }  \label{coord}
\end{equation}
. We see that (\ref{met}), (\ref{coord}) coincide with (\ref{c}) if the
constants are identified according to $a=Ce^{-CH_{h}}$, $B=2\tilde{\lambda}%
C^{-1}(1-2\kappa C)e^{CH_{h}}$.

\section{properties of solutions}

\subsection{Structure of spacetime}

Now we can make general conclusions about the structure of spacetime. Let us
restrict ourselves by the case $C=0$. It follows from (\ref{g}), (\ref{const}%
) that the curvature $R=-\frac{d^{2}g}{dx^{2}}$ is equal to 
\begin{equation}
R=\frac{4\lambda ^{2}}{\tilde{F}^{\prime }}[\frac{\omega (\tilde{F}-\tilde{F}%
_{h)}}{\tilde{F}^{\prime }}]^{\prime }\exp (\int^{\phi }d\tilde{\phi}\omega )
\label{r}
\end{equation}
Let, by definition, $\tilde{F}^{\prime }(\phi _{c})=0$, $x_{c}=x(\phi _{c})$%
. Near $\phi _{c}$ the function $\tilde{F}^{\prime }\propto \phi -\phi
_{c},x-x_{c}$ $\propto (\phi -\phi _{c})^{2}$ and, in general, $R\propto
(\phi -\phi _{c})^{-3}\propto (x-x_{c})^{-3/2}$. The exceptional case arises
when $\phi _{c}=\phi _{h}.$ Then the above expression in square brackets is
finite and $R\propto (\phi -\phi _{c})^{-1}\propto (x-x_{c})^{-1/2}$. Thus,
singularity becomes weaker but does not disappear. Such behavior, found
earlier for the RST model \cite{solod}, is inherent to any model under
consideration.

Thus, the metric possesses singularities in the points $\phi =\phi _{c}$
where $\tilde{F}^{\prime }=0.$ The spacetime splits into intervals between
zeros of $\tilde{F}^{\prime }$ which can be viewed as different sheets that
generalizes the corresponding feature of the RST model \cite{solod}. Within
each of them the function $\tilde{F}^{\prime }$ does not alter its sign, so
the function $\tilde{F}(\phi )$ is monotonic and the equation $\tilde{F}=%
\tilde{F}_{h\text{ }}$has only one root. Then, according to (\ref{g}), there
is only one horizon at $\phi =\phi _{h}$ on every sheet between any two
singularities with a finite value of $\phi $. In principle, it may happen
that on a sheet between infinity and a singularity nearest to it there
exists an additional horizon due to the factor $e^{-\psi }$ in which case
the coordinate $x$ calculated according to (\ref{g}) takes, generally
speaking, a finite value in this limit. To obtain the maximally extended
analytical continuation of spacetime, one is led to accept the possibility
of complex dilaton field values \cite{accel}. We will not, however, discuss
such possibilities further (a more detailed description of spacetime
structure will be done elsewhere\cite{prep}). For the RST model there exist
only two sheets but, depending on properties of the function $\tilde{F}(\phi
)$, the number of sheets in a general case can be made arbitrary. Any two
neighboring sheets are separated by the singularity located at $\phi =\phi
_{c}$.

As the singular points of the curvature represent zeros of the function $%
\tilde{F}^{\prime }$, the case under consideration admits solutions regular
everywhere, if the function $\tilde{F}^{\prime }$ does not turn into zero.
Let, for instance, $F=$ $e^{\omega \phi }+\delta \kappa \omega \phi $ where $%
\delta $ is a pure number and $\omega $ is constant. Then $\tilde{F}^{\prime
}=\omega [e^{\omega \phi }+\kappa (\delta -1)]$. If $\delta >1$, the
expression in square brackets changes its sign nowhere and does not tend to
zero at infinity. It follows from eq. (\ref{g}) that the coefficient $V$ for
such solutions is everywhere positive, so the action is well-defined. As far
as relationship between dilaton and coordinates is concerned, it follows
from (\ref{g}) that $\mu ^{\prime }=\omega +\omega \kappa (\delta
-1)e^{-\omega \phi }$, so $x=(2\lambda )^{-1}[\omega \phi -\kappa (\delta
-1)e^{-\omega \phi }]$ and $x\rightarrow \pm \infty $ when $\omega \phi
\rightarrow \pm \infty $. The metric function $g=1+e^{-\omega \phi }[(\delta
-1)\kappa \omega \phi -b]$, $b=[e^{\omega \phi }+\delta -1)\kappa \omega
\phi ]_{\phi =\phi _{h}}$ has one zero at $\phi =\phi _{h}$, $g\rightarrow 1$
at $\omega \phi \rightarrow \infty $ and $g\rightarrow -\infty $ at $\omega
\phi \rightarrow -\infty $. The curvature $R\simeq 4\lambda ^{2}[\kappa
(\delta -1)]^{-1}e^{-\left| \omega \phi \right| }\rightarrow 0$ when $\omega
\phi \rightarrow -\infty $ and $R\simeq -4\lambda ^{2}(\delta -1)\omega
\kappa \phi e^{-\omega \phi }\rightarrow 0$ when $\omega \phi \rightarrow
\infty $. Thus, not only the curvature is finite at both infinities but,
moreover, the spacetime is flat there. If $\kappa \rightarrow 0$ we return
to the string inspired dilaton gravity for which the singularity lies at $%
\omega \phi \rightarrow -\infty $. In this sense, it is quantum effects
which are responsible for removing the singularity. (On the other hand, one
can obtain the geometry regular everywhere already on a classical level due
to the choice, for example, $F=e^{\omega \phi }+K\omega \phi $ with $K>0$.)

Following general rules \cite{walk}, \cite{tom} we can draw the Penrose
diagram of this nonsingular spacetime starting from fundamental building
blocks. In so doing, the form of such a block depends crucially on whether
or not $f(x)=\int^{x}dyg(y)^{-1}$remains finite at the boundary: it is
triangle if $f$ is finite and square if $f$ diverges. In our case the
function $f$ can be calculated exactly from (\ref{g}): $f=(2\lambda
)^{-1}\ln (\tilde{F}-\tilde{F}_{h})$. Collecting all this information we
obtain the Penrose diagram which is depicted at Fig. \ref{pen1}. The
geodesic distance $\tau =\int dx(c-g)^{-1/2}$ where $c$ is a constant, so
timelike geodesics can reach either on plus or minus infinity only for an
infinite interval of time$.$ It is seen from Fig. \ref{pen1} that a horizon
is rather acceleration horizon than a true black hole one. The structure of
spacetime is similar to that of Rindler (with extension to the complete
Minkowski spacetime). In the limit $\omega \rightarrow 0$, $\delta
\rightarrow \infty $ with $\omega \delta =const$ this spacetime turns into
the Rindler one directly as it immediately follows from the above formulae
for the metric and coordinate. 
\begin{figure}[tbp]
\caption{Penrose diagram for a nonsingular spacetime with $\tilde{F}^{\prime
}\neq 0$ everywhere}
\label{pen1}
\end{figure}
It is worth noting that, as we shall see now, there exists also quite
another type of solutions regular everywhere and having a black hole
horizon: those for which $\tilde{F}^{\prime }\rightarrow 0$ at infinity in
such a way that cancellation of $\tilde{F}^{\prime }$ is compensated by $%
\exp (\int \omega d\phi )$, so that the whole expression (\ref{r}) remains
finite. Consider the following example. Let $\tilde{F}$ be monotonic
function of $\phi $ such that $\tilde{F}\simeq e^{\gamma _{\pm }\phi }$ when 
$\phi \rightarrow \pm \infty $ with $\gamma _{\pm }<0$. Let also the
function $\psi (\phi )=\int^{\phi }d\phi \omega $ have the asymptotics $\psi
\simeq \omega _{\pm }\phi +\psi _{\pm }$ when $\phi \rightarrow \pm \infty $%
. Then at both infinities the curvature (\ref{r}) behaves like $R\simeq
A_{\pm }\exp [(\omega _{\pm }-2\gamma _{\pm })\phi ]$ where $A_{\pm }$ are
constants. We choose $\omega _{+}-2\gamma _{+}<0$ and $\omega _{-}-2\gamma
_{-}>0,$ so $R\rightarrow 0$ when $\phi \rightarrow \pm \infty $. Let us
also choose, for definiteness, $\gamma _{-}=\omega _{-}$. Then $g\rightarrow
1$ when $\phi \rightarrow -\infty $ and $g\sim -e^{-\omega _{+}\phi
}\rightarrow -\infty $ when $\phi \rightarrow \infty $. The coordinate $%
x\simeq -2\lambda \left| \gamma _{-}\right| e^{-\psi _{-}}\phi \rightarrow
\infty $ at $\phi \rightarrow -\infty $ and $x\simeq -\left| \gamma
_{+}\right| (\gamma _{+}-\omega _{+})^{-1}e^{(\gamma _{+}-\omega _{+})\phi
}\rightarrow -\infty $ at $\phi \rightarrow \infty $ if $\gamma _{+}>\omega
_{+}$. Thus, the coordinate $-\infty <x<\infty $. All restrictions imposed
on parameters of the solutions read $\omega _{+}<2\gamma _{+}<\gamma _{+}<0$
and $\gamma _{-}=\omega _{-}<0$ and are self-consistent. With all these
properties, the Penrose diagrams looks like Fig. \ref{pen2} where the
horizontal lines represent not a singularity (as it would be for the
Schwarzschild metric) but regular spatial infinities which can be reached
along time-like geodesics only for a infinite proper time. The boundaries of
the spacetime under discussion are null- and time-like complete \cite{tom},
as $x\rightarrow \pm \infty $ and $\tau $ diverges. 

\begin{figure}[tbp]
\caption{Penrose diagram for a nonsingular black hole with $\tilde{F}%
^{\prime }\rightarrow 0$ at $\phi =\infty $.}
\label{pen2}
\end{figure}

As a matter of fact, constructing both types of diagrams relies only on the
asymptotic behavior of the functions $\tilde{F}$ and $\omega $, so they
describe the whole classes of spacetimes. Now we will show that our metrics (%
\ref{g}) contain also the third type of regular spacetimes - those with a
constant curvature. Indeed, let $\tilde{F}=\exp (\frac{1}{2}\int_{0}^{\phi
}\omega d\phi )$. Substituting this into (\ref{r}) we find after simple
manipulations that $R=8\lambda ^{2}\tilde{F}_{h}=const>0$, so spacetime in
question is of de Sitter type. This generalizes the observation made in \cite
{accel} for the particular choice $\omega =-2$ and $\kappa =0$.

The metrics (\ref{g}) corresponding to the choice $C=0$ possess the
following interesting property. If the functions $\omega (\phi )$ and $%
\tilde{F}(\phi )$, whatever their form would be, do not contain $\kappa $,
the metric also does not contain $\kappa $. Thus, within one-loop accuracy,
we obtained the whole classes of models for which a {\it classical} geometry
is the {\it exact} solution of field equations derived from {\it quantum}
Lagrangians.

\subsection{Solutions with a constant dilaton value}

Apart from solutions discussed above, there is one more class of them. It is
seen from eq. (\ref{13}) that this equation turns into identity when $\phi
=const\equiv \phi _{0}$. For such solutions field eq.(\ref{12}) gives us $%
U=-\kappa R$. Substituting it into eq.(\ref{9}) we have $R\tilde{F}^{\prime
}=0$ where we have taken into account that $\tilde{F}^{\prime }=u-\kappa
\omega $. This means that nontrivial solutions $(R\neq 0)$ exist only for
values of the dilaton field $\phi _{0}=\phi _{c}$. Let me recall that this
is just the point where the curvature for solutions described by eq.(\ref{g}
) diverges. This gives nontrivial interplay between two branches of
solutions, also found for the particular case of the RST model \cite{solod}:
the values of the dilaton field for constant dilaton solutions coincides
with the singularity of non-constant ones. In particular, it follows from
the contents of the present paragraph that the class of models under
discussion does not contain constant dilaton solutions with two horizons
found in \cite{zasl98}. It is not surprising since the latter solutions
exist only under the presence of an electromagnetic field which is now
absent.

\subsection{Thermodynamics}

Let us now return to the solutions with $\phi \neq const$ and discuss their
thermodynamic properties. From eqs. (\ref{c}), (\ref{const}) we obtain the
formula for the Hawking temperature $T_{H}$ which turns out surprisingly
simple: 
\begin{equation}
T_{H}=(4\pi )^{-1}(\frac{dg}{dx})_{x=x_{h}}=\frac{aB}{4\pi (1-2\kappa C)}
\label{temp}
\end{equation}
If the spacetime is not asymptotically flat, the choice of the Euclidean
time is ambiguous and this is reflected in the appearance of the constants $%
a $, $B$ in the metric (\ref{c}) and temperature. More interesting is,
however, the case when a spacetime is flat at infinity where there is a
clear definition of time and a distant observer measures a temperature in an
unambiguous way. Let us assume that $C=0$ and, for definiteness, the
asymptotically flat region corresponds to $\phi \rightarrow -\infty $ and $%
\tilde{F}\rightarrow \infty $, $\exp (-\int^{\phi }d\phi \omega )\rightarrow
0$ in such a way that their product is constant. Adjusting the limit of
integration, one can always achieve $\lim_{\phi \rightarrow -\infty }\tilde{F%
}\exp (-\int^{\phi }d\phi \omega )=a=1$ to have $g=1$ at infinity. Then it
follows from (\ref{g}), (\ref{temp}) that the temperature in this case 
\begin{equation}
T_{H}=(2\pi )^{-1}\lambda  \label{htemp}
\end{equation}
and does not acquire quantum corrections. Moreover, as this temperature is a
constant, it turns out that all horizons present in the solution have the
same temperature. Both properties generalize the similar feature of the RST
model \cite{solod}. In the latter case $\omega =const$, $\tilde{F}\propto
e^{\omega \phi }$.

It turns out that the general structure of the theories under consideration
enables one to relate entropy and energy of a system in a rather simple
manner. If the Hamiltonian constraint $T_{0}^{0}=0$ is taken into account
one can infer that the action takes the thermodynamic form $I=\beta E-S$.
This expression for the zero loop action can be obtained by direct
generalization of the procedure elaborated in \cite{action} for the
string-inspired dilaton gravity. Here $\beta =T_{H}^{-1}\sqrt{g}$ is the
inverse Tolman temperature on the boundary, the energy $E$ $=E_{o}-(2\pi
)^{-1}(\frac{\partial F}{\partial l})_{B}$ where $E_{o\text{ }}$is the
subtraction constant irrelevant for our purposes, $dl$ is the line element,
the index ''B'' indicates that the corresponding quantities are calculated
at the boundary. If quantum effects are taken into account the entropy
entering this formula is to be understood as $S$ $=S_{0}+S_{q}$ where zero
loop contribution $S_{0}=2F(\phi _{h})$ and that of Hawking radiation $%
S_{q}=2\kappa [\psi (\phi _{B})-\psi (\phi _{h})]$ \cite{qent} (we choose
the constant in the definition of entropy in such a way that $S_{q}=0$ in
the limit $\phi _{B}=\phi _{h}$ when there is no room for radiation). For
the total entropy we have $S=S_{0}+S_{q}=2[\tilde{F}(\phi _{h})+\kappa \psi
(\phi _{B})]$.

It follows directly from (\ref{g}) with $a=1$, $B=2\lambda $ that $E-E_{0}=-%
\frac{\lambda }{\pi }\exp (\frac{\psi _{B}}{2})\sqrt{F_{B}-S/2}$. The
general first law $\beta \delta E=\delta S$ then tells us that the Hawking
temperature $T_{H}=\lambda /2\pi $ does not acquire quantum correction in
agreement with the above formula (\ref{htemp}). It was pointed out in \cite
{solod} that not only the temperature but the energy and entropy as well do
not acquire quantum corrections in the RST model. Strictly speaking,
however, the physical meaning of this statement is not quite clear since it
refers to characteristics of a black hole itself and is based on subtracting
contributions of a hot gas \cite{solod} whereas in the microcanonical
ensemble approach the total energy (but not its constituents) should be
fixed. Meanwhile, it is seen from the above formula directly that if the
function $F(\phi )$ is such that it does not contain $\kappa $, the
dependence of the energy on total entropy retains the same form either in
the classical or quantum domain, so there are no quantum corrections for
characteristic of the whole system in this sense. The difference between
this situation and that discussed in \cite{solod} consists in that in the
RST model $F$ contains $\kappa $ but $V$ does not, whereas in our case the
situation is reverse according to eqs. (\ref{d=1}), (\ref{g}). It is worth
noting that the expression for $V$ can be rewritten as $V=-2\tilde{F}%
^{\prime }+2\kappa $. Therefore, although this coefficient may change its
sign it happens only after $\tilde{F}^{\prime }$ so does, i.e. beyond the
singularity. On the whole sheet with $\tilde{F}^{\prime }\leq 0$ the
quantity $V$ is positive as it should be for the action principle to be
well-defined.

It is also worth noting that if $\omega =const$ the formula for the entropy
of Hawking radiation can be written as $S_{q}=2\kappa \omega (\phi _{B}-\phi
_{h})$ that coincides with the entropy of hot gas in a flat space in the
container of the length $L=\left| \phi _{B}-\phi _{h}\right| =\lambda \left|
x_{B}-x_{h}\right| $ with a temperature $T=\lambda /2\pi $. This generalizes
the corresponding observation \cite{solod} for the RST model. One can say
that, in some sense, the entropy of thermal radiation in the backgrounds
under discussion does not acquire curvature corrections.

\subsection{Factorization}

It is instructive to look at the approach in question from somewhat another
view point. Eqs. (\ref{9}), (\ref{12}) lead to the relation 
\begin{equation}
R(u-\kappa \omega )=(2V-\omega u)\Box \phi +(V^{\prime }-\omega u^{\prime
})(\nabla \phi )^{2}  \label{28}
\end{equation}
If the potential $V$ satisfies eq. (\ref{g}), eq. (\ref{28}) takes the form 
\begin{equation}
(u-\kappa \omega )[R-\Box (\int d\phi \omega )]=0  \label{fact}
\end{equation}
Thus, our choice of relationship between action coefficients gives rise to
the important property of factorization, generalizing observation made for
the RST model in \cite{solod}. If $u=\kappa \omega $ we return to the
constant dilaton field solutions discussed above. Otherwise $R=\Box \omega $
and, according to (\ref{4}), $\omega =\psi ^{\prime }$ in agreement with (%
\ref{g}).

One can try to gain the factorization property without referring to eq. (\ref
{14}) and the form of $V$ following from it. Let us choose the functions,
entering the action of the model, in such a way that the coefficient at ($%
\nabla \phi )^{2}$ cancel and, besides, the coefficient at $\Box \phi $ be
proportional to that at curvature: 
\begin{equation}
V^{\prime }=\omega u^{\prime },\text{ }2V-\omega u=-\omega _{0}(u-\kappa
\omega )  \label{29}
\end{equation}
where $\omega _{0\text{ }}$is some constant. Then eq.(\ref{28}) turns into 
\begin{equation}
(u-\kappa \omega )(R+\omega _{0}\Box \phi )=0  \label{30}
\end{equation}
However, one can check directly that eq. (\ref{14}) is satisfied in this
case automatically, so we do not obtain new solutions. It is worth noting
that now $D\neq 1$ where $D$ is a constant entering (\ref{w}). There is also
the choice $V^{\prime }-\omega u^{\prime }=p(u-\kappa \omega )$, $2V-\omega
u=q(u-\kappa \omega )$ with arbitrary functions $p$,$q$ that ensures
factorization but we will not discuss this possibility further.

\section{comparison with known models}

We demonstrated above that the RST model enters our scheme as a particular
case. Here we show that the same is true for other known exactly solvable
models.

\subsection{BPP model}

This model \cite{bose} is characterized in our notations by 
\begin{equation}
F=e^{-2\phi }-2\kappa \phi \text{, }\tilde{F}=e^{-2\phi }\text{, }\omega =-2%
\text{, }\psi =2\phi \text{, }V=4e^{-2\phi }+2\kappa  \label{bpp}
\end{equation}
It is seen from (\ref{bpp}) immediately that (\ref{d=1}) is satisfied.
Substituting (\ref{bpp}) into (\ref{g}) we obtain 
\begin{equation}
g=1-e^{2\phi -2\phi _{h}}\text{, }-\lambda x=\phi  \label{gbpp}
\end{equation}
Thus, this model possesses rather unexpected feature: the metric has the
same form in terms of $\phi $ and $\phi _{h}$ (or $x$ and $x_{h}$) as its
classical counterpart! In other words, not only quantum corrections to the
Hawking temperature vanish but also so do quantum correction to the metric
itself. It is worth stressing that although this property sharply contrasts
with the explicit form of the metric listed in \cite{bose} there is no
contradiction here: authors of \cite{bose} consider solutions which are
radiationless at infinity and have a singular horizon (analog of the
Boulware state) whereas we deal with the Hartle-Hawking state that implies
that the stress-energy tensor of radiation does not vanish at infinity, an
event horizon being regular. In fact, as was explained in Sec. IIIA, any
model from our set for which $F=\tilde{F}(\phi )+\kappa \int d\phi \omega $
where $\tilde{F}(\phi )$ does not contain $\kappa $ will give the metric
function without quantum corrections as follows from (\ref{g}).

\subsection{Other models}

The model discussed by Michaud and Myers \cite{rob} is characterized by $%
F=e^{-2\phi }-\frac{\kappa }{2}(\alpha +\sum_{n=2}^{K}a_{n}\phi ^{n})$, $%
V=4e^{-2\phi }-\frac{\kappa }{2}(\beta +\sum_{n=2}^{K}b_{n}\phi ^{n-1})$, $%
\omega =-2$ where $b_{n}=-2na_{n}$, $\beta =4-2\alpha $.

For the Fabbri and Russo (FR) model \cite{fub} $F=\exp (-\frac{2\phi }{n}%
)+\kappa \frac{(1-2n)}{n}\phi $, $V=\frac{4}{n}\exp (-\frac{2\phi }{n}%
)+2\kappa \frac{(n-1)}{n}$, $\omega =-2$. When $n=1$ the RST result is
reproduced.

The model which interpolates between the RST and BPP ones is considered in 
\cite{cruz} (CN model). In this case $F=\exp (-2\phi )+2\kappa (a-1)\phi $, $%
V=4\exp (-2\phi )+2(1-2a)\kappa $, $\omega =-2$. It reduces to the RST model
when $a=\frac{1}{2}$ and to the BPP one when $a=0$.

It is easily seen that eq. (\ref{d=1}) is satisfied in all these cases.
Moreover, one can suggest, for instance, a new model which incorporates at
once features of the RST, FR, CN and BPP: $F=\exp (-\frac{2\phi }{n}%
)+2\kappa (a-1)$, $V=4\exp (-\frac{2\phi }{n})+2\kappa (1-2a)$, $\omega =-2$%
. When $a=2n^{-1}$ we return to the BPP, $n=1$ corresponds to the CN model.

For the exponential models \cite{cr2} $F=\phi $, $u=1$, $V=0$, $\omega
=const $. This case is described by eq. (\ref{w}) with $D\neq 1$.

\section{summary and outlook}

Thus, the models considered in the present paper have the following
properties:

1) they are exactly solvable in the sense that either the metric or dilaton
field are found in a closed form; 2) their geometry is static; 3) quantum
corrections to the Hawking temperature vanish; 4) for wide subsets of these
models one of the following properties holds: a) the relationship between
the total energy and the total entropy of the quantum finite size system is
the same as in the classical limit, b) the metric itself does not contain
quantum corrections and has the same form either in the classical or the
quantum domain; 5) there exists the special class of solutions with a
constant dilaton field $\phi =\phi _{c}$, the geometry of non-constant
dilaton solutions become singular in the point $\phi _{c}$; 6) all spacetime
for $\phi \neq \ const$ can be divided to separate sheets with one and only
one horizon on every sheet between two neighboring singularities with finite 
$\phi _{c}$ (plus, perhaps, additional horizons due to $\phi =\infty $ or $%
\phi =-\infty $), different sheets are glued in the singular points; all
horizons on different sheets share the same temperature; 7) there exists the
solution with one horizon and without singularities.

The most part of these properties is inherent to the RST model \cite{solod}.
It does not possess, however, the properties 4a) and 4b). Besides, in 6) the
number of sheets for the RST model is equal to two, whereas in general it
can be arbitrary. Thus, the RST model turns out to be only a representative
of a more wide class of models sharing common features. Moreover, we gained
also the qualitatively new property which was absent in the RST model: the
existence of quantum black holes without singularities whose metric is found
explicitly. In this respect the corresponding solutions resemble those in
string theory obtained in exact (non-perturbative) approach \cite{dijk}, 
\cite{perry}. On the other hand, regular black holes solutions found in the
present paper differ from similar ones in \cite{banks} in that solutions
discussed in \cite{banks} represent the extreme black holes whereas in our
case they are essentially nonextreme since their temperature is nonzero
constant.

The crucial point in which generalization of the RST model is performed
consists in that we do not specify the form of the action coefficients and
only impose one restriction (\ref{15}) on them, so instead of particular
models we operate with whole classes of them. In this respect our approach
is similar to that of \cite{kaz} but, in contrast to it, we appeal directly
to properties of field equations and do not rely on the general structure of
the nonlinear sigma model from which the dilaton-gravity action can be
obtained. In so doing, we only assumed (i) the local connection between two
unknown functions $\psi $ and $\phi $; (ii) cancellations of coefficients at
first and second derivatives of $\phi $ in eq.(\ref{13}) that saved us from
the trouble to solve a generic complicated differential equation for $\phi
(x)$ and enabled us to introduce variables in terms of which the field
equations are greatly simplified.

It is of interest to generalize the elaborated approach to theories with
higher derivatives (in particular, to generalize exact solutions for black
holes found at the classical level \cite{od}), additional scalar and gauge
fields, interactions between black holes spacetimes and shock waves, etc.
The problem deserving separate attention is detailed desription and
classification of different types of spacetime structure of considered
quantum black holes similarily to what has been done for classical dilaton
black holes \cite{accel}, \cite{jose}. Of special interest is the issue of
regular black holes including their formation by gravitation collapse
starting form the vacuum. It would be temting to derive general criteria for
Lagrangians admitting nonsingular black hole solutions either in the
nonextreme or extreme case and select among them those with exact solutions.
Besides the issues connected with black hole physics, exactly solvable
models can be useful for the analysis of conceptual problems in different
schemes of quantization of dilaton gravity theories \cite{jakiw}.

\section{acknowledgments}

I am grateful to Sergey Solodukhin for stimulating remarks and to J. V. Cruz
for drawing my attention to a number of papers on exactly solvable models in
dilaton gravity. This work is supported by International Science Education
Program (ISEP), grant No. QSU082068.





%
%
%
%


\begin{references}
\bibitem{callan}  C. G. Callan, Giddings, J. A. Harvey, and A. Strominger,
Phys. Rev. D{\bf \ 45} , R1005 (1992); T.Banks, A. Dabholkar, M. R. Douglas,
and M. O. 'Loughlin, Phys. Rev.{\bf \ }D{\bf \ 45}, 3607 (1992).

\bibitem{rst}  J. G. Russo, L. Susskind, and L. Thorlacius, Phys. Rev. D{\bf %
\ 46}, 3444 (1992); Phys. Rev. D{\bf \ 47}, 533 (1992).

\bibitem{solod}  S. N. Solodukhin, Phys. Rev. D{\bf \ 53}, 824 (1996).

\bibitem{bil}  A. Bilal and C. G. Callan, Nucl. Phys. {\bf B} 394, 73 (1993).

\bibitem{alw}  S. P. de Alwis, Phys. Rev. D {\bf 46}, 5429 (1992).

\bibitem{kaz}  Y. Kazama, Y. Satoh, and A. Tsuichiya, Phys. Rev. D {\bf 51},
4265 (1995).

\bibitem{rob}  G. Michaud and R. C. Myers, Two-Dimensional Dilaton Black
Holes, gr-qc/9508063.

\bibitem{fub}  A. Fabbri and J. G. Russo, Phys. Rev. D {\bf 53}, 6995 (1995).

\bibitem{bose}  S. Bose, L. Parker, and Y. Peleg, Phys. Rev. D {\bf 52},
3512 (1995).

\bibitem{cruz}  J. Cruz and J. Navarro-Salas, Phys. Lett. B 375, 47 (1996).

\bibitem{cr2}  J. Cruz, J. Navarro-Salas, M. Navarro and C. F. Talavera,
Phys. Lett. B {\bf 402}, 270 (1997).

\bibitem{polyakov}  A. M. Polyakov, Phys. Lett. B {\bf 103}, 207 (1981).

\bibitem{israel}  V. P. Frolov, W. Israel, and S. N. Solodukhin, Phys. Rev. D%
{\bf \ 54,} 2732 (1996).

\bibitem{string}  G. Mandal, A. Sengupta, and S. Wadia, Mod. Phys. Lett. A%
{\bf \ 6}, 1685 (1991); E. Witten, Phys. Rev. D{\bf \ 44}, 314 (1991).

\bibitem{accel}  R. Balbinot and A. Fabbri, Class. Quant. Grav. {\bf 13},
2457 (1996)

\bibitem{prep}  O. B. Zaslavskii, in preparation.

\bibitem{walk}  M. Walker, J. Math. Phys. {\bf 11} 2280 (1970).

\bibitem{tom}  T. Kl\"{o}sch and T. Strobl, Class. Quant. Grav. {\bf 13},
2395 (1996).

\bibitem{zasl98}  O. B. Zaslavskii, Phys. Lett.{\bf \ }B{\bf \ 424}, 271
(1998) (hep-th/9802117).

\bibitem{action}  V. P. Frolov, Phys. Rev. D{\bf \ 46}, 5383 (1992);
G.W.Gibbons and M.J.Perry, Int.J.Mod.Phys.{\bf \ }D{\bf \ 1} 335 (1992).

\bibitem{qent}  R. C. Myers, Phys. Rev. {\bf D} 50, 6412 (1994); D. V.
Fursaev and S. N. Solodukhin, Phys. Rev. D {\bf 52} (1995) 2133.

\bibitem{dijk}  R. Dijkkgraaf, H. Verlinde, and E. Verlinde, Nucl. Phys. B%
{\bf \ 371}, 269 (1992).

\bibitem{perry}  M. J. Perry and E. Teo, Phys. Rev. Lett. {\bf 70}, 2669
(1993).

\bibitem{banks}  T. Banks and M. O' Loughlin, Phys. Rev. D {\bf 48}, 698
(1993).

\bibitem{od}  E. Elizalde, P. Fosibla-Vela, S. Naftulin, and S. Odintsov,
Phys. Lett. {\bf B} 352, 237 (1995).

\bibitem{jose}  J. P. S. Lemos and P. S\'{a}, Phys. Rev. D {\bf 49}, 2897
(1994), Erratum Phys. Rev. D {\bf 51}, 5967 (1995).

\bibitem{jakiw}  E.Benedict, R. Jackiw, and H.-J.Lee, Phys. Rev. D {\bf 54},
6213 (1996); R. Jakiw and C. Teitelboim, in {\it Quantum Theory of Gravity},
edited by S. Christensen (Hilger, Bristol, UK, 1984).
\end{references}
\end{document}